\documentclass[sigconf,nonacm]{acmart}

\usepackage{tabularx}
\usepackage{booktabs}
\usepackage{threeparttable} 
\usepackage{makecell}       

\AtBeginDocument{%
  }

\copyrightyear{2024}
\acmYear{2024}
\setcopyright{acmlicensed}\acmConference[SIGCSE Virtual 2024]{Proceedings of the 2024 ACM Virtual Global Computing Education Conference V. 1}{December 5--8, 2024}{Virtual Event, NC, USA}
\acmBooktitle{Proceedings of the 2024 ACM Virtual Global Computing Education Conference V. 1 (SIGCSE Virtual 2024), December 5--8, 2024, Virtual Event, NC, USA}
\acmDOI{10.1145/3649165.3690125}
\acmISBN{979-8-4007-0598-4/24/12}

\begin{document}

\title[Integrating Natural Language Prompting Tasks]{Integrating Natural Language Prompting Tasks in Introductory Programming Courses}

\author{Chris Kerslake}
\orcid{0000-0001-9860-941X}
\affiliation{%
  \institution{Simon Fraser University}
  \city{Burnaby}
  \state{British Columbia}
  \country{Canada}
}
\email{chris.kerslake@sfu.ca}

\author{Paul Denny}
\orcid{0000-0002-5150-9806}
\affiliation{
  \institution{University of Auckland}
  \city{Auckland}
  \country{New Zealand}
}
\email{paul@cs.auckland.ac.nz}

\author{David H. Smith IV}
\orcid{0000-0002-6572-4347}
\affiliation{
  \institution{University of Illinois}
  \city{Urbana, IL}
  \country{USA}
}
\email{dhsmith2@illinois.edu}

\author{James Prather}
\orcid{0000-0003-2807-6042}  
\affiliation{  
\institution{Abilene Christian University}
\city{Abilene, Texas}  
\country{USA}  
}  
\email{james.prather@acu.edu}

\author{Juho Leinonen}
\orcid{0000-0001-6829-9449}  
\affiliation{  
\institution{Aalto University}
\city{Espoo}  
\country{Finland}  
}  
\email{juho.2.leinonen@aalto.fi}

\author{Andrew Luxton-Reilly}
\orcid{0000-0001-8269-2909}  
\affiliation{  
\institution{University of Auckland}  
\city{Auckland}  
\country{New Zealand}  
}  
\email{a.luxton-reilly@auckland.ac.nz}

\author{Stephen MacNeil}
\affiliation{%
\institution{Temple University}
\city{Philadelphia}  
\state{PA}  
\country{US}}  
\email{stephen.macneil@temple.edu}
\orcid{0009-0001-1597-3513}

\renewcommand{\shortauthors}{Kerslake, Denny, Smith, Prather, Leinonen, Luxton-Reilly, MacNeil}

\begin{abstract}

Introductory programming courses often emphasize mastering syntax and basic constructs before progressing to more complex and interesting programs. This bottom-up approach can be frustrating for novices, shifting the focus away from problem solving and potentially making computing less appealing to a broad range of students. 
The rise of generative AI for code production could partially address these issues by fostering new skills via interaction with AI models, including constructing high-level prompts and evaluating code that is automatically generated.
In this experience report, we explore the inclusion of two prompt-focused activities in an introductory course, implemented across four labs in a six-week module. The first requires students to solve computational problems by writing natural language prompts, emphasizing problem-solving over syntax. The second involves students crafting prompts to generate code equivalent to provided fragments, to foster an understanding of the relationship between prompts and code.
Most of the students in the course had reported finding programming difficult to learn, often citing frustrations with syntax and debugging.  We found that self-reported difficulty with learning programming had a strong inverse relationship with performance on traditional programming assessments such as tests and projects, as expected.  However, performance on the natural language tasks was less strongly related to self-reported difficulty, suggesting they may target different skills.  Learning how to communicate with AI coding models is becoming an important skill, and natural language prompting tasks may appeal to a broad range of students.

\end{abstract}

\ccsdesc[500]{Social and professional topics~Computing education}

\keywords{introductory programming, LLM, CS1, natural language prompting, prompt engineering, explain in plain English, EiPE}



\maketitle

\section{Introduction}

Introductory programming courses traditionally focus on teaching students to write code in a bottom-up fashion, starting with mastering syntax and basic constructs, and gradually progressing to building more complex and interesting programs.  Problem solving is often considered the most engaging aspect of programming, but the difficulties novices face with syntax, errors, and low-level code-layout issues can detract from this focus and cause frustration making computing courses less appealing to a diverse range of students \cite{becker2019compiler}.  Moreover, the widespread use of generative AI for producing code raises questions about when and how these tools should be introduced in introductory courses \cite{porter_learn_2024}.

Large language models (LLMs) have shown impressive capabilities for solving computational tasks when provided with appropriate natural language prompts \cite{denny2023conversing, prather2023robots, finnie2022robots}.  Thus, there are two emerging skills that students need to develop in this generative AI era.  The first is being able to construct clear, unambiguous prompts to express desired solutions to computational tasks, and the second is understanding and evaluating the code generated by these models, to verify that it is indeed solving the intended tasks.  While students may now be developing these skills independently of the curriculum, there is value in explicitly teaching students how to construct effective prompts \cite{denny2024computing}.  

In this experience report we describe the inclusion of two kinds of prompt-focused tasks alongside traditional activities in an introductory programming course.  Both kinds of tasks involve students writing only natural language prompts for an LLM.  The first task involves students solving computational tasks by writing prompts to generate code.  This is a very authentic activity in today's landscape, with a focus on problem-solving rather than on code syntax.  The second task involves showing students a code fragment and asking them to demonstrate their understanding of the code by crafting a prompt that generates equivalent code.  The two tasks are complementary, as the first allows students to explore the relationship between computational problems and high-level prompts, and the second allows students to explore the relationship between high-level prompts and code.  We are interested in understanding if the skills needed to solve these prompting tasks are distinct from those needed to be successful with traditional programming tasks.

We collected self-reported data from students on how difficult they found learning programming and why they found it difficult.  We categorize students based on self-reported difficulty and then compare their performance on traditional programming assessments with their performance on the natural language prompting tasks.  We also explore their perceptions of seeing these tasks integrated alongside more traditional tasks and present examples of some of the prompts that students created.  Our evaluation is guided by the following overarching question: \emph{How successful are students at natural language prompting tasks compared to more traditional programming assessments, and how does this vary by self-reported difficulty of learning to program?}

\section{Related Work}

Developing the ability to comprehend and communicate the behaviour of code,
though always considered an important set of skills for novice programmers to
develop~\cite{azad2020lessons, whalley_australasian_2006}, is necessary to work
effectively with LLMs~\cite{smith_code_2024, denny2024explaining,
prather2023robots,sarsa2022automatic}. This requires 1) the ability to describe
the requirements of a problem with sufficient detail for it to be implemented in
code and 2) the ability to understand the purpose of code. The former is needed
as poorly constructed prompts are less likely to generate desired
solutions~\cite{denny2023conversing} and the latter is needed to evaluate the
code that is generated~\cite{prather2023robots, dakhel2023github}.

\subsection{Teaching Prompting}\label{sec:teaching_prompting}

To develop student skills in expressing problems effectively, instructors have
explored various tasks to provide students practice with
prompting in formative contexts. \citet{denny2024prompt} introduced ``Prompt
Problems'', an activity where students are shown visual representations that
illustrate specific instances of a task, asked to infer the general problem from
the specific cases shown, and then provide a prompt that generates code that
performs the task. The generated code is then graded based on an
instructor-defined suite of test cases to determine if the generated code is
functionally correct~\cite{denny2024prompt}. \citet{nguyen2024beginning}
evaluated similar activities where students were shown input-output pairs, asked
to infer the task being performed, create a prompt that generates code with that
functionality, and then evaluate if the generated code is correct. Their
findings highlight that many students struggle to form successful prompts,
understand generated code, and have poor mental models of generative AI, which hinders
their ability to form effective prompting strategies~\cite{nguyen2024beginning}.

\subsection{Explain in Plain English Questions}\label{sec:eipe_lit}

Prompting an LLM to generate code has some similarity to the `Explain in plain
English' (EiPE) task, which is commonly used to assess student comprehension of
code~\cite{murphy2012explain}. In these activities, students are given a code
sample and asked to describe the purpose of the
code~\cite{whalley_australasian_2006}. EiPE activities are designed to focus on
the code's high-level (abstract) purpose rather than the details of
implementation (i.e., the mechanics of how it achieves the purpose).
Unfortunately, novice programmers often struggle to describe the purpose of code
in this way~\cite{bonar1983uncovering} suggesting greater emphasis should be
placed on comprehension tasks in CS1. However, due to the difficulty of
developing rubrics~\cite{fowler2021how} and autograders for such
questions~\cite{azad2020lessons, azad2020strategies,fowler2021autograding,
li2023wrong, hsu2021attitudes}, their adoption for use in formative settings
has been limited.

Responses to both Prompt Problems and EiPE questions require a natural language
problem specification. In the case of EiPE, this takes the form of students
inferring the purpose of the code by reading it, in effect, reverse engineering
the prompt that could be used to generate the given code. Using the grading
approach of ~\citet{smith_code_2024}, the success of a student's prompt can be
judged based on whether or not an LLM can successfully generate the desired
code. Additionally, this approach both eases the difficulty of developing
autograded EiPE questions and provides students with feedback via the generated
code and test cases~\cite{denny2024explaining}.

\begin{table*}[ht]
\caption{List of questions and their descriptions from the four labs.}
\label{tab:questions}
\centering
\begin{tabularx}{\textwidth}{ccclX}
\hline
\textbf{Activity} & \textbf{Lab} & \textbf{Question} & \textbf{Task} & \textbf{Description} \\
\hline
EiPE & 8 & 1 & FindSumBetween & calculates the sum between a `low' and `high' value \\
& & 2 & CountEvensInArray & counts the number of even values in an array \\
& & 3 & LastZero & finds the position of the last occurrence of zero in an array \\
& & 4 & SumPositiveValues & calculates the sum of all positive values in an array \\
\hline
Prompt Problems & 9 & 1 & Average & replaces each value in an array with the average of those values \\
& & 2 & Sum & calculates the sum of all of the even numbers in an array \\
& & 3 & Find & finds the index position of the last occurrence of zero in an array \\
\hline
EiPE & 10 & 1 & ReverseString & reverses a string in place \\
& & 2 & FindSumOfGivenRow & calculates the sum of all values on a row in a 2D array \\
& & 3 & DoesStringContainVowel & checks if a string contains a vowel \\
& & 4 & DoesStringContainSubstring & checks if a string contains a substring \\
\hline
Prompt Problems & 12 & 1 & TwoQueens & determines if two queens attack each other on a chessboard \\
& & 2 & FullQueens & determines if eight queens are placed without attacking each other \\
& & 3 & LeafEater & calculates how many leaves a bug eats as it moves along a branch \\
\hline
\end{tabularx}
\end{table*}

\section{Approach}

\begin{figure}
\centering
  \includegraphics[width=.85\linewidth]{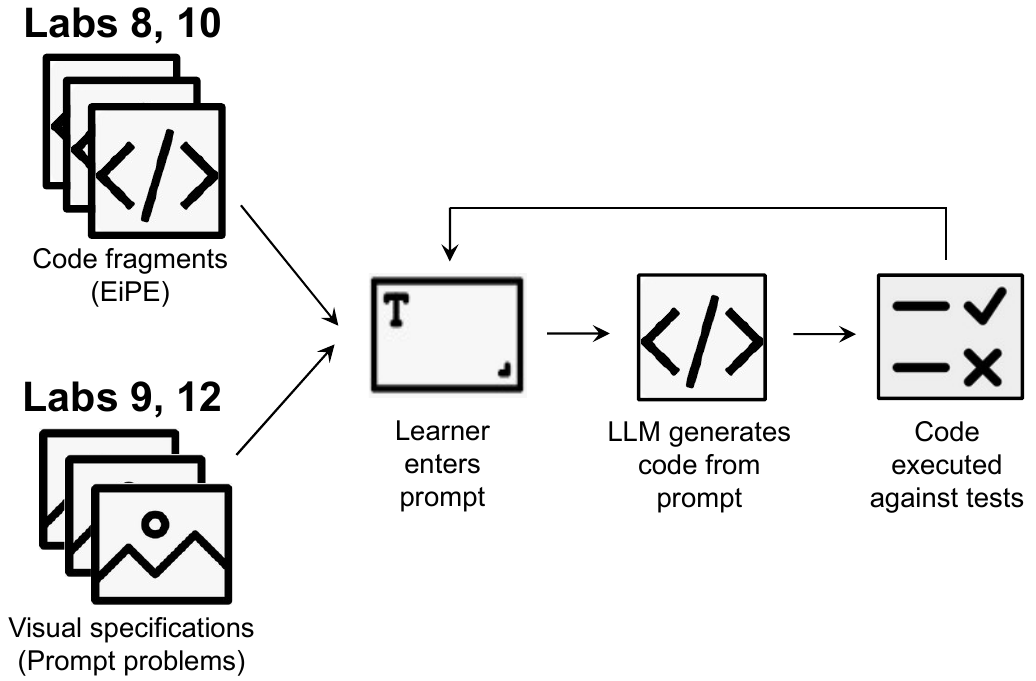}
  \caption{Natural language prompting tasks requiring the learner to enter a prompt in response to either a code fragment (EiPE) or a visual problem specification (Prompt Problem).  The prompt is sent to an LLM, and the resulting code is programmatically evaluated.  The EiPE and Prompt Problems were interleaved across Labs 8, 9, 10 and 12.} 
  \label{fig:schematic}
  \Description{Natural language programming tasks explored in this work}
\end{figure}

In this paper, we incorporated two kinds of natural-language prompting tasks (see Figure \ref{fig:schematic}) into a course covering traditional CS1 topics and explored student perceptions and success with these tasks in comparison to more traditional programming-focused assessments (both invigilated and non-invigilated).  The course was taught over a 12-week semester, although we focused on integrating these new tasks during the second half of the course (weeks 7--12).  This timing allowed students to reflect on their progress throughout the first half of the course (weeks 1--6), including reporting how difficult they found programming to learn, before any exposure to the natural-language tasks.

\subsection{Course Context}

The course, \{\emph{Anon}\}, is taught at \{\emph{Institution Anonymized}\} which is a large research university in \{\emph{Country Anonymized}\}.  The course is designed for engineering students and is structured into two modules, each spanning 6 weeks.  The first module covers typical CS1 topics, including variables, arithmetic, arrays (vectors), functions, control flow, and basic algorithms using MATLAB.  The second module introduces the C programming language and reinforces the concepts covered in the first half of the course.  

\subsubsection{Student Participants}
Following approval by the university's human ethics committee, data was collected from 861 students enrolled in the 
course.  Most students have no formal programming experience, but some enter with prior experience based on their choices from high school.

\subsection{Assessments and Reflection}

There are three large, invigilated assessments in the course (one test for each module and a final exam), which account for 56\% of the final grade in the course.  The course also includes weekly programming-focused lab sessions (24\% weighting) and one project (10\% each) for each of the two modules, all of which are non-invigilated.  

After the first module covering MATLAB, students were asked to reflect on their experience learning programming and respond with the extent to which they agree with the statement:

\begin{itemize}
    \item \emph{I find programming difficult.}
\end{itemize}  

Responses were collected using a standard 5-step Likert-response scale from ``Strongly disagree'' (SD) to ``Strongly agree'' (SA).


In addition to this question at the beginning of the second module in the course, each lab session included an optional post-lab survey that invited students to comment on any aspect of the lab. 

\subsection{Natural Language Tasks}

We incorporated two kinds of natural language tasks across four of the six weekly lab sessions in the second half of the course.  Eight `Explain in Plain English' (EiPE) tasks were included in Labs 8 and 10, and six Prompt Problem tasks were included in Labs 9 and 12.  Table \ref{tab:questions} summarises these 14 problems.

\subsubsection{Explain in Plain English (EiPE) Questions}
\begin{figure}[htbp]
    \centering
    \includegraphics[width=0.9\columnwidth]{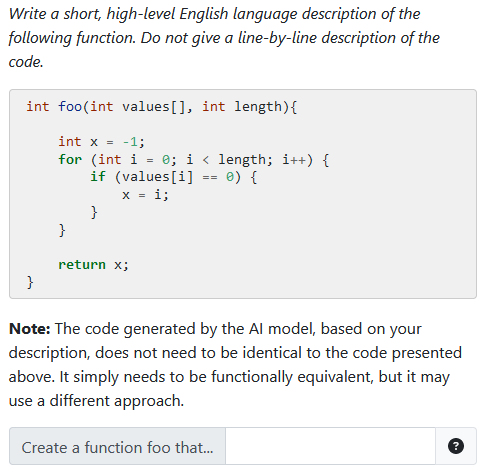}
    \caption{An example EiPE problem. 
    The ability to highlight and copy the code is disabled to dissuade students from copying it into ChatGPT.}
    \label{fig:pl-indexlastzero}
  \Description{Example of an EiPE problem}
\end{figure}

In the tradition of EiPE questions \cite{fowler2021how, murphy2012explain}, for each task, students were presented with a single function and instructions indicating that they should describe the function in plain English (see Figure~\ref{fig:pl-indexlastzero}). To prevent giving away the
code's purpose, the variables were replaced with generic names and each function was named \texttt{foo}. Tasks were delivered using PrairieLearn, an open-source online assessment platform \cite{west2015prairielearn}. After the student description was submitted, a prompt to generate a solution meeting the description was passed to GPT-3.5.  The code was evaluated against a set of test cases and then displayed to students with the results of the tests.

\subsubsection{Prompt Problems}
\begin{figure}
\centering
  \includegraphics[width=.87\linewidth]{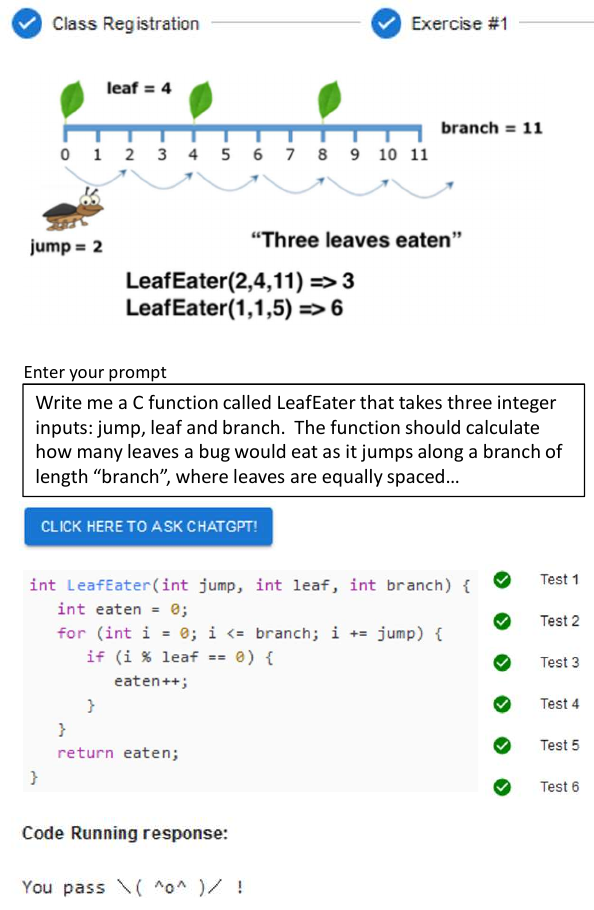}
  \caption{Interface layout for a Prompt Problem showing the LeafEater task (Lab 12, Question 3).} 
  \label{fig:promptly}
  \Description{Example of promptly}
\end{figure}

A Prompt Problem consists of a visual presentation of a computational task, to which a student must craft an LLM prompt to generate code that solves the task.  In our course, we used a custom tool similar to the ones described by Denny et al. \cite{denny2024prompt} and by Nguyen et al. \cite{nguyen2024beginning}.  When viewing a Prompt Problem in our tool (see Figure \ref{fig:promptly}), the student sees a visual representation of the problem and enters their prompt as plain text.  When their prompt is submitted, the verbatim text is sent to an LLM along with some additional prompting to guide the model to produce only code and no additional explanatory text.  The generated code is executed automatically, and the test case output is shown.


\section{Findings}
We organize the findings using students' self-reported answers to the survey question ``I find programming difficult'' collected at the end of Lab 7 since this provides an intuitive grouping of like-minded and potentially like-skilled students.  We started with 861 students, and removed any who did not complete each of the required labs (8, 9, and 10), or who missed any of the course exams or projects, which resulted in 726 students for analysis.

\subsection{Summary of Difficulty Data}
A common performance pattern emerged in the course based on students' reported difficulty with programming. As shown in Table \ref{tab:llm-task-success-attempts}, students who did not report programming as difficult (Disagree or Strongly disagree) on average scored the highest on all exams and projects.  In contrast, students who reported that programming was difficult (Agreed or Strongly agreed) scored the lowest on average on all exams and projects.

\begin{table*}[ht]

\caption{LLM task attempts and success by student group.  EiPE (Labs 8 \& 10), Prompt Problems (Lab 9*). Difficult (Diff.) was self-reported by students during Lab 7 via a 5-point Likert scale question asking if they agreed that programming was difficult. Tries are the average number of attempts for each group. First indicates the percentage of students who completed a task on the first attempt, and Final is the percentage of students who completed the task after one or more attempts.  The Projects column lists the combined averages for each groups' uninvigilated MATLAB and C programming projects.}
\label{tab:llm-task-success-attempts}
\begin{threeparttable}
\centering
\def\arraystretch{1.5}
\begin{tabularx}{.8288\textwidth}{|c|c|c|c|c|c|c|c|c|c|c|c|}
\hline
\multicolumn{2}{|c|}{} & \multicolumn{3}{|c|}{\textbf{EiPE Problems}} & \multicolumn{3}{|c|}{\textbf{Prompt Problems}*} & \multicolumn{3}{|c|}{\textbf{Invigilated Exams}} & \textbf{Projects}\\
\hline
\textbf{Diff.} & \textbf{N} & \textbf{Tries} & \textbf{First} & \textbf{Final} & \textbf{Tries} & \textbf{First} & \textbf{Final} & \textbf{MATLAB} & \textbf{C} & \textbf{Final} & \textbf{MATLAB+C}\\
\hline
D/SD** &  68 & 1.68 & 67.5\% & 99.6\% & 3.26 & 55.9\% & 100.0\% & 72.1\% & 82.4\% & 87.0\% & 93.8\%\\
\hline
N    & 208 & 1.82 & 63.0\% & 99.7\% & 3.61 & 53.7\% & 100.0\% & 57.2\% & 77.0\% & 83.9\% & 92.4\%\\
\hline
A    & 311 & 1.88 & 60.0\% & 99.4\% & 3.68 & 52.5\% & 100.0\% & 48.0\% & 69.8\% & 78.3\% & 87.3\%\\
\hline
SA   & 139 & 1.89 & 59.9\% & 98.7\% & 4.32 & 45.8\% &  99.8\% & 42.8\% & 66.0\% & 73.1\% & 82.8\%\\
\hline
\end{tabularx}
\begin{tablenotes}\footnotesize
\item[*] Unlike Labs 8, 9, and 10, Lab 12 was excluded because it was optional and less than 9\% of students participated.
\item[**] Only 13 students selected Strongly disagree, so they were merged with the Disagree students to create D/SD.
\end{tablenotes}
\end{threeparttable}
\end{table*}

\subsection{Student Attempt Success}
\begin{figure}
\centering
  \includegraphics[width=.9\linewidth]{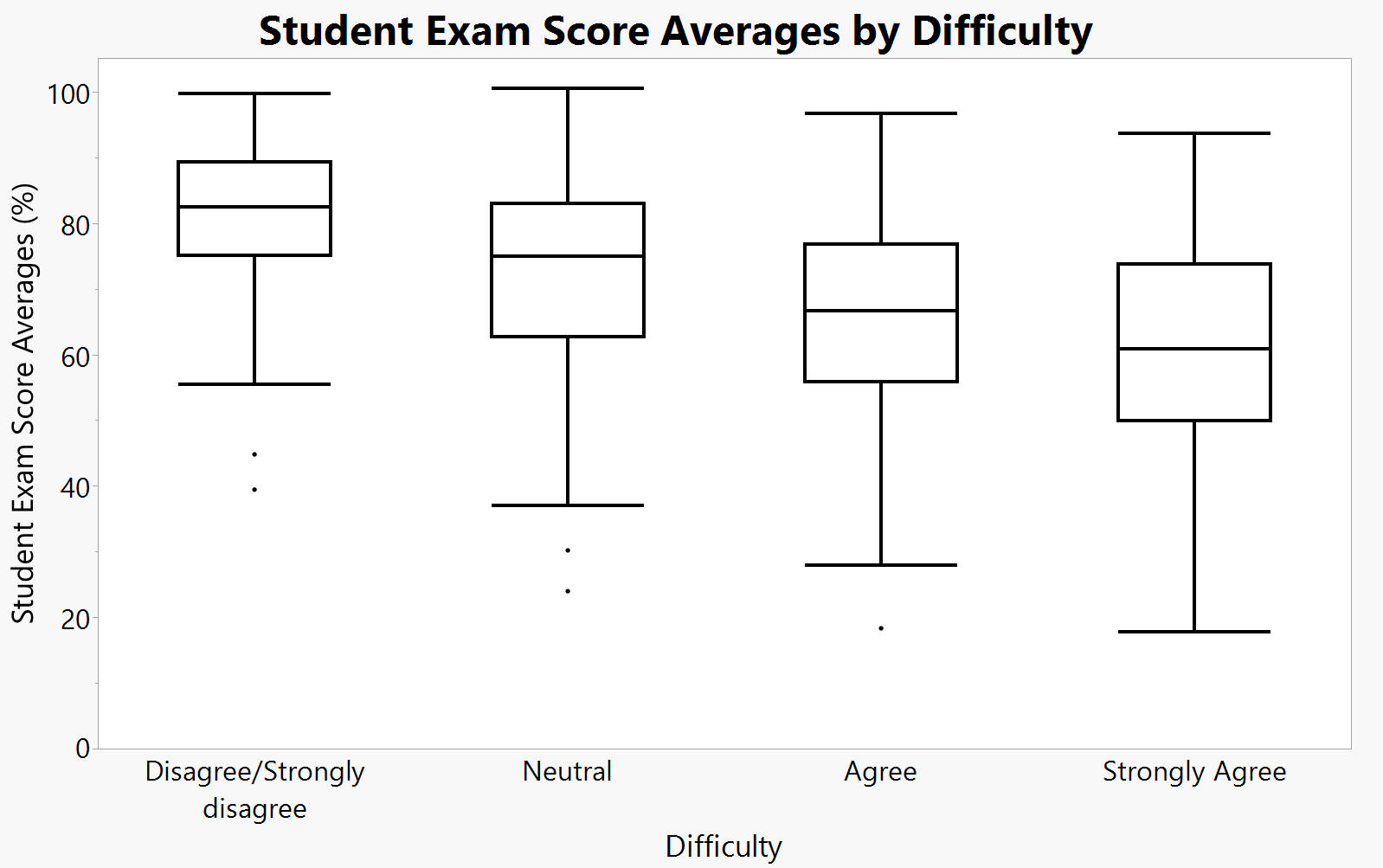}
  \caption{
  A comparison of student exam score averages for three invigilated exams grouped by difficulty.} 
  \label{fig:box-exams-by-diff}
  \Description{Student Exam Score Averages by Difficulty. Compares student exam score averages for three invigilated exams, MATLAB, C, and final exam, grouped by difficulty.}
\end{figure}

\begin{figure}
\centering
  \includegraphics[width=.9\linewidth]{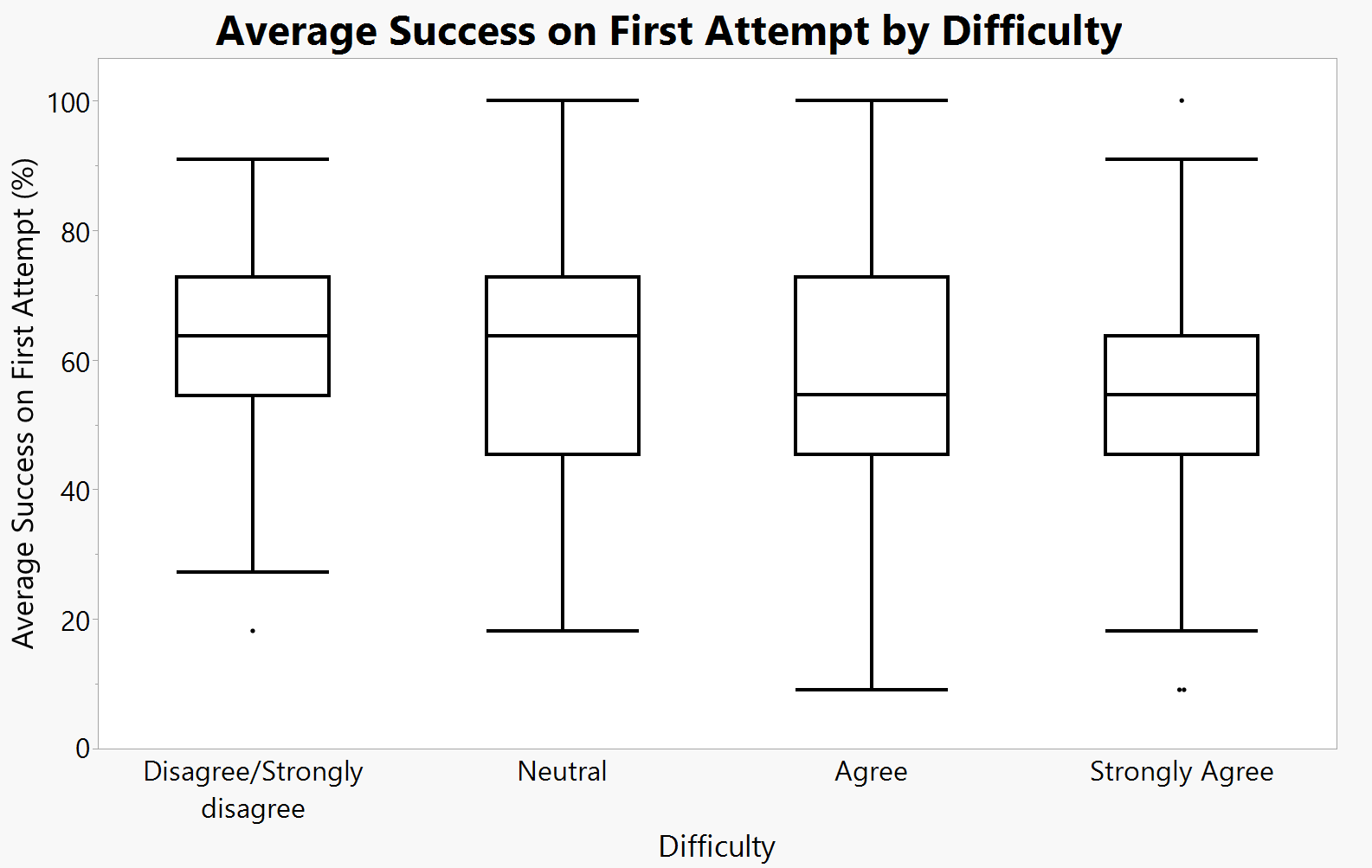}
  \caption{
  A comparison of students' average success rate on their first attempt to solve the LLM-related questions. The optional Lab 12 was excluded due to low participation.
  } 
  \label{fig:box-first-success-by-difficulty}
  \Description{Average Success on First Attempt by Difficulty.  Compares students' average success rate on their first attempt to solve each LLM-related question for labs 8, 9, and 10.  Lab 12 was excluded due to it being optional and having low participation.}
\end{figure}

For both the EiPE and Prompt Problems, we calculated the success of each student's first attempt to solve the problem (First) as well as whether they eventually were successful for each question, along with the number of attempts (Tries in Table \ref{tab:llm-task-success-attempts}).  The first-attempt calculations were viewed as similar to how students would engage with problems during exams, so we compared the two to see whether students performed similarly on each or not.  The final attempt (Final in Table \ref{tab:llm-task-success-attempts}) was calculated for whether students completed the tasks, similar to the course projects.  The number of attempts for each type of task was calculated to determine how many attempts each of the different difficulty groups spent on the two types of problems.
As detailed in Table \ref{tab:llm-task-success-attempts}, all student groups were much closer in performance on the LLM problems (Figure \ref{fig:box-first-success-by-difficulty}) than they were in their invigilated exam performance (Figure \ref{fig:box-exams-by-diff}).


\subsection{Examples of Code Explanations}
The code descriptions shown in Table \ref{tab:reverse-string-prompts-small} were provided by students during Lab 10 in response to code that reversed a string (ReverseString).  Students provided varied prompts, but interestingly, students at both ends of the difficulty scale provided similar prompts (e.g., students 19 and 1).  Additionally, some students relied on direct citation of the code as they saw it, as shown by student 105.  Also of interest is that student 77 used the term ``backwards'' which did not result in the LLM successfully generating functionally equivalent code; in contrast to how a human might understand and interpret their meaning.  However, student 13's response also used ``backwards'', but incorrectly.  The implications are that a human reviewer might be able to understand the nuanced use of the term ``backwards'', but in this specific situation, the LLM could not.

\begin{table}[ht]
\caption{Student EiPE explanation examples for Lab 10's ``ReverseString'' question, and whether they passed test cases.}
\label{tab:reverse-string-prompts-small}
\centering
\def\arraystretch{1.5}
\begin{tabular}{ccl}
\hline
\textbf{Diff.} & \textbf{Pass} & \textbf{Code Explanation (Student No.)}\\
\hline
\makecell{D/SD} &  Yes & \makecell{``reverses a string'' (19)}\\
\hline
N    & Yes & \makecell{``takes one string as input and loops till length\\of the string - 1 and replaces str i with str of j\\ and replaces str of j with str of i which is\\called a char temp, and increases i and\\ decreases j index" (105)}\\
\hline
A    &  Yes & \makecell{``reverses the input string array'' (1)}\\
\hline
D/SD &  No  & \makecell{``takes a string and turns it backwards'' (77)}\\
\hline
N    &  No  & \makecell{``writes words in a sentence backwards'' (13)}\\
\hline
SA   & No & \makecell{``converts a character input array to an\\ output array of its ascii values'' (106)}\\
\hline
\end{tabular}
\end{table}

\subsection{Student Comments on Programming}
\begin{figure}
\centering
  \includegraphics[width=.9\linewidth]{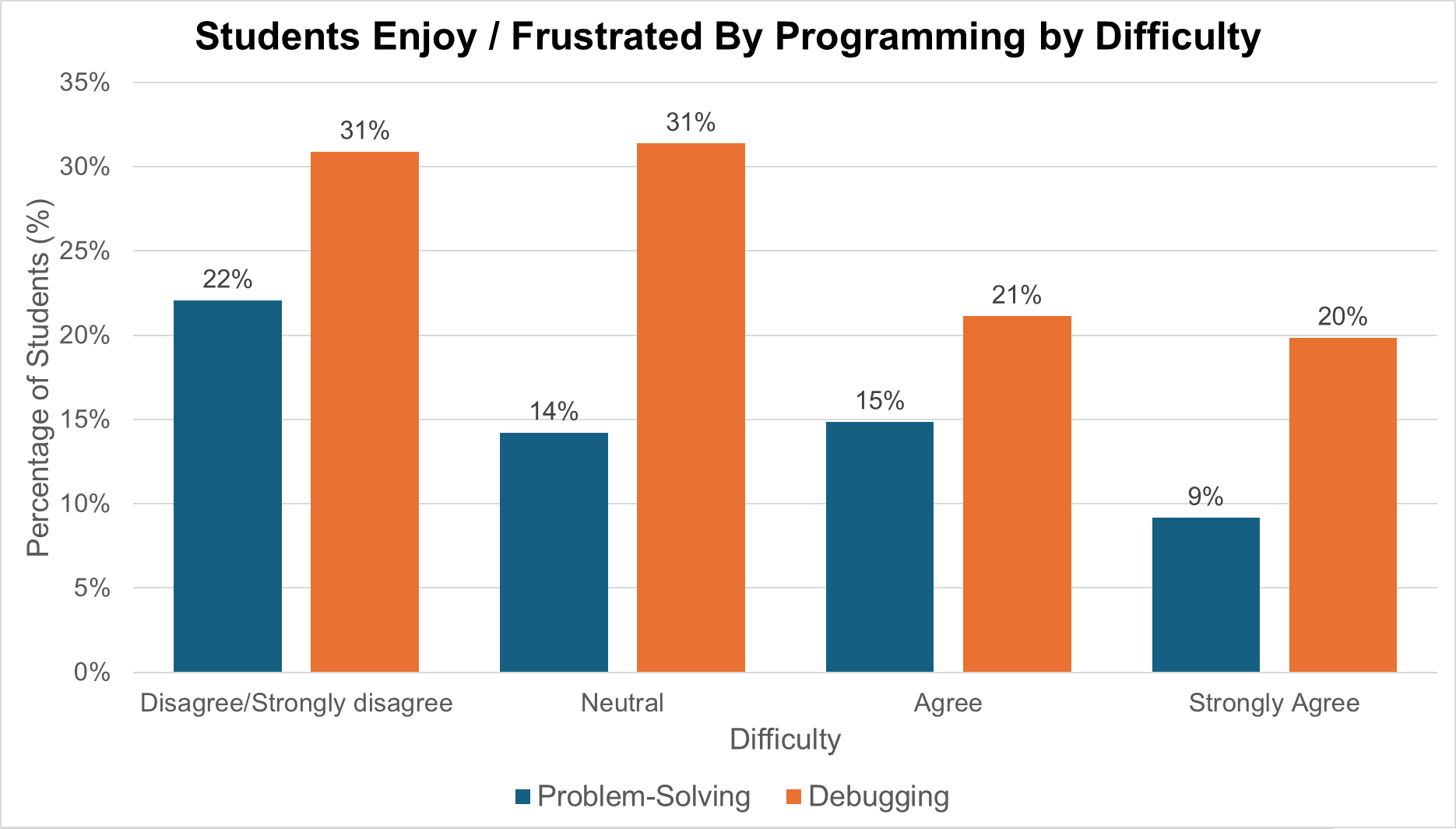}
  \caption{
  Percentage of students in each difficulty group who reported `problem-solving' for enjoyment and `debugging' for frustration.}
  \label{fig:hist-enjoy_frustration_reasons}
  \Description{Reasons Students Reported Enjoying and Being Frustrated By Programming by Difficulty.  Percentage of students in each difficulty group who reported 'problem-solving' for enjoyment and 'debugging' for frustration during a survey at the end of Lab 7.}
\end{figure}

At the end of Lab 7, students were asked to complete a feedback survey about the course.  One question asked students what they enjoyed most and found most frustrating about programming.  
The most common enjoyment was problem-solving, and the most common frustration was debugging.  As shown in Figure \ref{fig:hist-enjoy_frustration_reasons}, problem-solving was the most common enjoyment for students who reported that they enjoyed programming (D/SD; 22\%) and the least reported enjoyment for students who reported that programming was difficult (SA; 9\%).  Additionally, debugging was reported as the most common source of frustration. Unexpectedly, students who reported the least difficulty (D/SD \& N) reported it the most frequently (31\%), although the other groups still reported it ~20\% of the time.

\subsection{Student Perceptions}\label{sec:student-perceptions}
At the end of each LLM-related lab, students were asked to provide their thoughts on the lab and the activities.  Several students provided feedback after the EiPE questions at the end of Labs 8 and 10, shown below.  Overall, students were predominantly positive about the experience, with a single student calling the task ``gimmicky'' and less effective at assessing their performance than traditional code-writing tasks.
\begin{itemize}
    \item \emph{Positive: ``I thought the code comprehension task was good, because it encourages understanding of code logic without the pressure of having to write code or debug. It also helps to improve my ability to communicate what a piece of code does by forcing me to write clear and concise explanations of code that can be easily understood.'' (402, Strongly Agree; Difficult)}
    \item \emph{Positive: ``I found the starting of the code comprehension task really difficult. This was because as I was doing it I was over complicating it each time. However with more attempts and practice it became easier. I really liked this task as it helped me on how to understand and condense my understanding with basic concepts in coding.'' (329, Agree; Difficult)}
    \item \emph{Positive: ``[EiPE problems] also enlightened me on the different solutions that could be available to solve a specific task'' (510, Neutral)}
    \item \emph{Negative: ``Dislike the [EiPE] tasks. Feels gimmicky. It's to me as if I saw this new AI-powered tool, and I'm using it just because I learnt about it [rather than for the tool's merits].  I don't think it proves any student performance, and the code-writing based lab tasks prove our understanding much more.'' (350, Disagree/Strongly disagree; Easy)}
\end{itemize}


\section{Discussion and Takeaways}
As noted by Porter \& Zingaro \cite{porter_learn_2024}, the introduction of LLMs suggests re-evaluating how we teach introductory programming courses.  Additionally, with increased class sizes, the need to automate learning at scale highlights the importance of identifying techniques and tools that can help students assess and improve their code comprehension abilities.  However, they also note that LLMs do not replace the need to understand what the code does when generated, nor does it remove the need for being able to explain the problem sufficiently to produce the desired results.  The tasks we have analyzed in this work -- Prompt Problems and EiPE questions -- aim to teach students these skills explicitly via generative AI, and provide automated feedback that can support their use at scale.

Recent work has shown that the introduction of generative AI into computing classrooms is negatively impacting student programming skills like code writing and debugging. Jost et al. found a significant negative correlation between increased use of LLMs for coding tasks and lower critical thinking skills as well as a decrease in final grades \cite{jovst2024impact}. Prather et al. found that generative AI code completion tools like Copilot and ChatGPT can benefit some students who are already confident in their programming abilities, but that it can be directly harmful to the programming problem solving ability of students who are not \cite{prather2024widening}. Students who already have difficulty with programming, like many of the ones we examined in this study, seem to be the most poised to over-rely on generative AI \cite{margulieux_self-regulation_2024} and its harms could be compounded upon them.

Our findings suggest that natural language tasks could `bridge the gap' between students who struggle with traditional assessments and those who do not, which could engage a broader range of students and possibly address the harmful impacts of generative AI on this group. Additionally, most students found the natural language programming tasks positive, indicating it is not useless for those who do not struggle. Most students required multiple attempts to craft a prompt that correctly solved their given task, similar to prior work~\cite{nguyen2024beginning,denny2024prompt}, which provides additional support for explicitly teaching students `prompt engineering.' Previous work has found that this kind of iterative learning while solving Prompt Problems could support the development of metacognitive skills in novice programmers \cite{prather2024interactions}, which is possibly why these tasks can help address the potential negative impacts of generative AI. 


There are many possible explanations for the students' performance being more similar on the natural language tasks than traditional code writing tasks. One is that success on the natural language tasks does not rely on mastery of low-level syntax, which can be hard for novices to get right~\cite{denny2012syntax,leinonen2019exploring}. Another is that the natural language tasks are simply easier, and therefore less likely to differentiate between students who are more or less confident with programming.  However, typical success rates on the first attempts at both problem types (usually in the 50\% to 60\% range) suggest that the tasks are not trivial to solve.  Finally, a further possible explanation is that these tasks target different skills altogether compared to traditional assessments. In traditional code writing tasks, students can, for example, tinker their way to a solution~\cite{lehtinen2023automated}, unable to explain the code they have written~\cite{lehtinen2021students}, which signals that they do not understand it. In the natural language tasks, the student's goal is at a higher level of abstraction -- they do not need to think about how to write the code to solve the problem but instead describe how the code works at a high level (EiPE), or describe the functionality of the code in natural language (Prompt Problems).

The finding that most students found the natural language tasks positive is similar to findings in prior work. Nguyen et al.~\cite{nguyen2024beginning} found that most students would use a natural language programming tool again if it were available, and Denny et al.~\cite{denny2024prompt} reported that most students found similar tasks educational.

\section{Conclusion}

In this work, we describe our experience including two kinds of natural language prompting tasks alongside traditional assessments in an introductory programming course. 
We observed a weak relationship between performance on these tasks and more traditional programming assessments, suggesting that these new tasks may assess a different set of competencies.  We also collected self-reported data from students on the difficulty they experienced learning programming.  Interestingly, self-reported difficulty was very strongly related to performance on tests, exams, and programming projects, as would be expected, but this was not the case for the natural language tasks.  In other words, students with less prior experience or those who find traditional programming challenging appear to perform relatively well on these tasks, potentially reducing the advantage experienced students typically have. We also found that students appreciated these types of tasks and recognized the importance of learning about AI and its applications in programming.  


\balance

\bibliographystyle{ACM-Reference-Format}
\bibliography{SIGCSE-Virtual-2024}


\begin{thebibliography}{30}


\ifx \showCODEN    \undefined \def \showCODEN     #1{\unskip}     \fi
\ifx \showDOI      \undefined \def \showDOI       #1{#1}\fi
\ifx \showISBNx    \undefined \def \showISBNx     #1{\unskip}     \fi
\ifx \showISBNxiii \undefined \def \showISBNxiii  #1{\unskip}     \fi
\ifx \showISSN     \undefined \def \showISSN      #1{\unskip}     \fi
\ifx \showLCCN     \undefined \def \showLCCN      #1{\unskip}     \fi
\ifx \shownote     \undefined \def \shownote      #1{#1}          \fi
\ifx \showarticletitle \undefined \def \showarticletitle #1{#1}   \fi
\ifx \showURL      \undefined \def \showURL       {\relax}        \fi
\providecommand\bibfield[2]{#2}
\providecommand\bibinfo[2]{#2}
\providecommand\natexlab[1]{#1}
\providecommand\showeprint[2][]{arXiv:#2}

\bibitem[Azad(2020)]%
        {azad2020lessons}
\bibfield{author}{\bibinfo{person}{Sushmita Azad}.} \bibinfo{year}{2020}\natexlab{}.
\newblock \emph{\bibinfo{title}{Lessons learnt developing and deploying grading mechanisms for EiPE code-reading questions in CS1 classes}}.
\newblock \bibinfo{thesistype}{Ph.\,D. Dissertation}. \bibinfo{school}{University of Illinois at Urbana-Champaign}.
\newblock


\bibitem[Azad et~al\mbox{.}(2020)]%
        {azad2020strategies}
\bibfield{author}{\bibinfo{person}{Sushmita Azad}, \bibinfo{person}{Binglin Chen}, \bibinfo{person}{Maxwell Fowler}, \bibinfo{person}{Matthew West}, {and} \bibinfo{person}{Craig Zilles}.} \bibinfo{year}{2020}\natexlab{}.
\newblock \showarticletitle{Strategies for {Deploying} {Unreliable} {AI} {Graders} in {High}-{Transparency} {High}-{Stakes} {Exams}}. In \bibinfo{booktitle}{\emph{21st {International} {Conference}, {AIED} 2020, {Ifrane}, {Morocco}, {July} 6–10, 2020, {Proceedings}, {Part} {I}}}, \bibfield{editor}{\bibinfo{person}{Ig~Ibert Bittencourt}, \bibinfo{person}{Mutlu Cukurova}, \bibinfo{person}{Kasia Muldner}, \bibinfo{person}{Rose Luckin}, {and} \bibinfo{person}{Eva Millán}} (Eds.), Vol.~\bibinfo{volume}{LNAI 12163}. \bibinfo{publisher}{Springer Cham}, \bibinfo{address}{Cham, Switzerland}, \bibinfo{pages}{16--28}.
\newblock
\showISBNx{978-3-030-52237-7}
\urldef\tempurl%
\url{https://doi.org/10.1007/978-3-030-52237-7_2}
\showDOI{\tempurl}


\bibitem[Becker et~al\mbox{.}(2019)]%
        {becker2019compiler}
\bibfield{author}{\bibinfo{person}{Brett~A. Becker}, \bibinfo{person}{Paul Denny}, \bibinfo{person}{Raymond Pettit}, \bibinfo{person}{Durell Bouchard}, \bibinfo{person}{Dennis~J. Bouvier}, \bibinfo{person}{Brian Harrington}, \bibinfo{person}{Amir Kamil}, \bibinfo{person}{Amey Karkare}, \bibinfo{person}{Chris McDonald}, \bibinfo{person}{Peter-Michael Osera}, \bibinfo{person}{Janice~L. Pearce}, {and} \bibinfo{person}{James Prather}.} \bibinfo{year}{2019}\natexlab{}.
\newblock \showarticletitle{Compiler Error Messages Considered Unhelpful: The Landscape of Text-Based Programming Error Message Research}. In \bibinfo{booktitle}{\emph{Proceedings of the Working Group Reports on Innovation and Technology in Computer Science Education}} (Aberdeen, Scotland, UK) \emph{(\bibinfo{series}{ITiCSE-WGR '19})}. \bibinfo{publisher}{Association for Computing Machinery}, \bibinfo{address}{New York, NY, USA}, \bibinfo{pages}{177–210}.
\newblock
\showISBNx{9781450375672}
\urldef\tempurl%
\url{https://doi.org/10.1145/3344429.3372508}
\showDOI{\tempurl}


\bibitem[Bonar and Soloway(1983)]%
        {bonar1983uncovering}
\bibfield{author}{\bibinfo{person}{Jeffrey Bonar} {and} \bibinfo{person}{Elliot Soloway}.} \bibinfo{year}{1983}\natexlab{}.
\newblock \showarticletitle{Uncovering principles of novice programming}. In \bibinfo{booktitle}{\emph{Proceedings of the 10th {ACM} {SIGACT}-{SIGPLAN} symposium on {Principles} of programming languages}} \emph{(\bibinfo{series}{{POPL} '83})}. \bibinfo{publisher}{Association for Computing Machinery}, \bibinfo{address}{New York, NY, USA}, \bibinfo{pages}{10--13}.
\newblock
\showISBNx{978-0-89791-090-3}
\urldef\tempurl%
\url{https://doi.org/10.1145/567067.567069}
\showDOI{\tempurl}


\bibitem[Dakhel et~al\mbox{.}(2023)]%
        {dakhel2023github}
\bibfield{author}{\bibinfo{person}{Arghavan~Moradi Dakhel}, \bibinfo{person}{Vahid Majdinasab}, \bibinfo{person}{Amin Nikanjam}, \bibinfo{person}{Foutse Khomh}, \bibinfo{person}{Michel~C. Desmarais}, {and} \bibinfo{person}{Zhen Ming~(Jack) Jiang}.} \bibinfo{year}{2023}\natexlab{}.
\newblock \showarticletitle{{GitHub} {Copilot} {AI} pair programmer: {Asset} or {Liability}?}
\newblock \bibinfo{journal}{\emph{Journal of Systems and Software}}  \bibinfo{volume}{203} (\bibinfo{date}{Sept.} \bibinfo{year}{2023}), \bibinfo{pages}{111734}.
\newblock
\showISSN{0164-1212}
\urldef\tempurl%
\url{https://doi.org/10.1016/j.jss.2023.111734}
\showDOI{\tempurl}


\bibitem[Denny et~al\mbox{.}(2023)]%
        {denny2023conversing}
\bibfield{author}{\bibinfo{person}{Paul Denny}, \bibinfo{person}{Viraj Kumar}, {and} \bibinfo{person}{Nasser Giacaman}.} \bibinfo{year}{2023}\natexlab{}.
\newblock \showarticletitle{Conversing with {Copilot}: {Exploring} {Prompt} {Engineering} for {Solving} {CS1} {Problems} {Using} {Natural} {Language}}. In \bibinfo{booktitle}{\emph{Proceedings of the 54th {ACM} {Technical} {Symposium} on {Computer} {Science} {Education} {V}. 1}} \emph{(\bibinfo{series}{{SIGCSE} 2023})}. \bibinfo{publisher}{Association for Computing Machinery}, \bibinfo{address}{New York, NY, USA}, \bibinfo{pages}{1136--1142}.
\newblock
\showISBNx{978-1-4503-9431-4}
\urldef\tempurl%
\url{https://doi.org/10.1145/3545945.3569823}
\showDOI{\tempurl}


\bibitem[Denny et~al\mbox{.}(2024a)]%
        {denny2024prompt}
\bibfield{author}{\bibinfo{person}{Paul Denny}, \bibinfo{person}{Juho Leinonen}, \bibinfo{person}{James Prather}, \bibinfo{person}{Andrew Luxton-Reilly}, \bibinfo{person}{Thezyrie Amarouche}, \bibinfo{person}{Brett~A. Becker}, {and} \bibinfo{person}{Brent~N. Reeves}.} \bibinfo{year}{2024}\natexlab{a}.
\newblock \showarticletitle{Prompt {Problems}: {A} {New} {Programming} {Exercise} for the {Generative} {AI} {Era}}. In \bibinfo{booktitle}{\emph{Proceedings of the 55th {ACM} {Technical} {Symposium} on {Computer} {Science} {Education} {V}. 1}} \emph{(\bibinfo{series}{{SIGCSE} 2024})}. \bibinfo{publisher}{Association for Computing Machinery}, \bibinfo{address}{New York, NY, USA}, \bibinfo{pages}{296--302}.
\newblock
\showISBNx{9798400704239}
\urldef\tempurl%
\url{https://doi.org/10.1145/3626252.3630909}
\showDOI{\tempurl}


\bibitem[Denny et~al\mbox{.}(2012)]%
        {denny2012syntax}
\bibfield{author}{\bibinfo{person}{Paul Denny}, \bibinfo{person}{Andrew Luxton-Reilly}, {and} \bibinfo{person}{Ewan Tempero}.} \bibinfo{year}{2012}\natexlab{}.
\newblock \showarticletitle{All syntax errors are not equal}. In \bibinfo{booktitle}{\emph{Proceedings of the 17th ACM Annual Conference on Innovation and Technology in Computer Science Education}} \emph{(\bibinfo{series}{ITiCSE '12})}. \bibinfo{publisher}{Association for Computing Machinery}, \bibinfo{address}{New York, NY, USA}, \bibinfo{pages}{75–80}.
\newblock
\showISBNx{9781450312462}
\urldef\tempurl%
\url{https://doi.org/10.1145/2325296.2325318}
\showDOI{\tempurl}


\bibitem[Denny et~al\mbox{.}(2024b)]%
        {denny2024computing}
\bibfield{author}{\bibinfo{person}{Paul Denny}, \bibinfo{person}{James Prather}, \bibinfo{person}{Brett~A. Becker}, \bibinfo{person}{James Finnie-Ansley}, \bibinfo{person}{Arto Hellas}, \bibinfo{person}{Juho Leinonen}, \bibinfo{person}{Andrew Luxton-Reilly}, \bibinfo{person}{Brent~N. Reeves}, \bibinfo{person}{Eddie~Antonio Santos}, {and} \bibinfo{person}{Sami Sarsa}.} \bibinfo{year}{2024}\natexlab{b}.
\newblock \showarticletitle{Computing Education in the Era of Generative AI}.
\newblock \bibinfo{journal}{\emph{Commun. ACM}} \bibinfo{volume}{67}, \bibinfo{number}{2} (\bibinfo{date}{Jan} \bibinfo{year}{2024}), \bibinfo{pages}{56–67}.
\newblock
\showISSN{0001-0782}
\urldef\tempurl%
\url{https://doi.org/10.1145/3624720}
\showDOI{\tempurl}


\bibitem[Denny et~al\mbox{.}(2024c)]%
        {denny2024explaining}
\bibfield{author}{\bibinfo{person}{Paul Denny}, \bibinfo{person}{David~H. Smith}, \bibinfo{person}{Max Fowler}, \bibinfo{person}{James Prather}, \bibinfo{person}{Brett~A. Becker}, {and} \bibinfo{person}{Juho Leinonen}.} \bibinfo{year}{2024}\natexlab{c}.
\newblock \showarticletitle{Explaining {Code} with a {Purpose}: {An} {Integrated} {Approach} for {Developing} {Code} {Comprehension} and {Prompting} {Skills}}. In \bibinfo{booktitle}{\emph{Proceedings of the 2024 on {Innovation} and {Technology} in {Computer} {Science} {Education} {V}. 1}} \emph{(\bibinfo{series}{{ITiCSE} 2024})}. \bibinfo{publisher}{Association for Computing Machinery}, \bibinfo{address}{New York, NY, USA}, \bibinfo{pages}{283--289}.
\newblock
\showISBNx{9798400706004}
\urldef\tempurl%
\url{https://doi.org/10.1145/3649217.3653587}
\showDOI{\tempurl}


\bibitem[Finnie-Ansley et~al\mbox{.}(2022)]%
        {finnie2022robots}
\bibfield{author}{\bibinfo{person}{James Finnie-Ansley}, \bibinfo{person}{Paul Denny}, \bibinfo{person}{Brett~A. Becker}, \bibinfo{person}{Andrew Luxton-Reilly}, {and} \bibinfo{person}{James Prather}.} \bibinfo{year}{2022}\natexlab{}.
\newblock \showarticletitle{The Robots Are Coming: Exploring the Implications of OpenAI Codex on Introductory Programming}. In \bibinfo{booktitle}{\emph{Proceedings of the 24th Australasian Computing Education Conference}} (Virtual Event, Australia) \emph{(\bibinfo{series}{ACE '22})}. \bibinfo{publisher}{Association for Computing Machinery}, \bibinfo{address}{New York, NY, USA}, \bibinfo{pages}{10–19}.
\newblock
\showISBNx{9781450396431}
\urldef\tempurl%
\url{https://doi.org/10.1145/3511861.3511863}
\showDOI{\tempurl}


\bibitem[Fowler et~al\mbox{.}(2021b)]%
        {fowler2021autograding}
\bibfield{author}{\bibinfo{person}{Max Fowler}, \bibinfo{person}{Binglin Chen}, \bibinfo{person}{Sushmita Azad}, \bibinfo{person}{Matthew West}, {and} \bibinfo{person}{Craig Zilles}.} \bibinfo{year}{2021}\natexlab{b}.
\newblock \showarticletitle{Autograding "{Explain} in {Plain} {English}" questions using {NLP}}. In \bibinfo{booktitle}{\emph{Proceedings of the 52nd {ACM} {Technical} {Symposium} on {Computer} {Science} {Education}}} \emph{(\bibinfo{series}{{SIGCSE} '21})}. \bibinfo{publisher}{Association for Computing Machinery}, \bibinfo{address}{New York, NY, USA}, \bibinfo{pages}{1163--1169}.
\newblock
\showISBNx{978-1-4503-8062-1}
\urldef\tempurl%
\url{https://doi.org/10.1145/3408877.3432539}
\showDOI{\tempurl}


\bibitem[Fowler et~al\mbox{.}(2021a)]%
        {fowler2021how}
\bibfield{author}{\bibinfo{person}{Max Fowler}, \bibinfo{person}{Binglin Chen}, {and} \bibinfo{person}{Craig Zilles}.} \bibinfo{year}{2021}\natexlab{a}.
\newblock \showarticletitle{How should we ‘Explain in plain English’? Voices from the Community}. In \bibinfo{booktitle}{\emph{Proceedings of the 17th ACM Conference on International Computing Education Research}} (Virtual Event, USA) \emph{(\bibinfo{series}{ICER 2021})}. \bibinfo{publisher}{Association for Computing Machinery}, \bibinfo{address}{New York, NY, USA}, \bibinfo{pages}{69–80}.
\newblock
\showISBNx{9781450383264}
\urldef\tempurl%
\url{https://doi.org/10.1145/3446871.3469738}
\showDOI{\tempurl}


\bibitem[Hsu et~al\mbox{.}(2021)]%
        {hsu2021attitudes}
\bibfield{author}{\bibinfo{person}{Silas Hsu}, \bibinfo{person}{Tiffany~Wenting Li}, \bibinfo{person}{Zhilin Zhang}, \bibinfo{person}{Max Fowler}, \bibinfo{person}{Craig Zilles}, {and} \bibinfo{person}{Karrie Karahalios}.} \bibinfo{year}{2021}\natexlab{}.
\newblock \showarticletitle{Attitudes {Surrounding} an {Imperfect} {AI} {Autograder}}. In \bibinfo{booktitle}{\emph{Proceedings of the 2021 {CHI} {Conference} on {Human} {Factors} in {Computing} {Systems}}} \emph{(\bibinfo{series}{{CHI} '21})}. \bibinfo{publisher}{Association for Computing Machinery}, \bibinfo{address}{New York, NY, USA}, \bibinfo{pages}{1--15}.
\newblock
\showISBNx{978-1-4503-8096-6}
\urldef\tempurl%
\url{https://doi.org/10.1145/3411764.3445424}
\showDOI{\tempurl}


\bibitem[Jo{\v{s}}t et~al\mbox{.}(2024)]%
        {jovst2024impact}
\bibfield{author}{\bibinfo{person}{Gregor Jo{\v{s}}t}, \bibinfo{person}{Viktor Taneski}, {and} \bibinfo{person}{Sa{\v{s}}o Karakati{\v{c}}}.} \bibinfo{year}{2024}\natexlab{}.
\newblock \showarticletitle{The Impact of Large Language Models on Programming Education and Student Learning Outcomes}.
\newblock \bibinfo{journal}{\emph{Applied Sciences}} \bibinfo{volume}{14}, \bibinfo{number}{10} (\bibinfo{year}{2024}), \bibinfo{pages}{4115}.
\newblock


\bibitem[Lehtinen et~al\mbox{.}(2023)]%
        {lehtinen2023automated}
\bibfield{author}{\bibinfo{person}{Teemu Lehtinen}, \bibinfo{person}{Lassi Haaranen}, {and} \bibinfo{person}{Juho Leinonen}.} \bibinfo{year}{2023}\natexlab{}.
\newblock \showarticletitle{Automated Questionnaires About Students’ JavaScript Programs: Towards Gauging Novice Programming Processes}. In \bibinfo{booktitle}{\emph{Proceedings of the 25th Australasian Computing Education Conference}} (Melbourne, VIC, Australia) \emph{(\bibinfo{series}{ACE '23})}. \bibinfo{publisher}{Association for Computing Machinery}, \bibinfo{address}{New York, NY, USA}, \bibinfo{pages}{49–58}.
\newblock
\showISBNx{9781450399418}
\urldef\tempurl%
\url{https://doi.org/10.1145/3576123.3576129}
\showDOI{\tempurl}


\bibitem[Lehtinen et~al\mbox{.}(2021)]%
        {lehtinen2021students}
\bibfield{author}{\bibinfo{person}{Teemu Lehtinen}, \bibinfo{person}{Aleksi Lukkarinen}, {and} \bibinfo{person}{Lassi Haaranen}.} \bibinfo{year}{2021}\natexlab{}.
\newblock \showarticletitle{Students Struggle to Explain Their Own Program Code}. In \bibinfo{booktitle}{\emph{Proceedings of the 26th ACM Conference on Innovation and Technology in Computer Science Education V. 1}} (Virtual Event, Germany) \emph{(\bibinfo{series}{ITiCSE '21})}. \bibinfo{publisher}{Association for Computing Machinery}, \bibinfo{address}{New York, NY, USA}, \bibinfo{pages}{206–212}.
\newblock
\showISBNx{9781450382144}
\urldef\tempurl%
\url{https://doi.org/10.1145/3430665.3456322}
\showDOI{\tempurl}


\bibitem[Leinonen et~al\mbox{.}(2019)]%
        {leinonen2019exploring}
\bibfield{author}{\bibinfo{person}{Antti Leinonen}, \bibinfo{person}{Henrik Nygren}, \bibinfo{person}{Nea Pirttinen}, \bibinfo{person}{Arto Hellas}, {and} \bibinfo{person}{Juho Leinonen}.} \bibinfo{year}{2019}\natexlab{}.
\newblock \showarticletitle{Exploring the Applicability of Simple Syntax Writing Practice for Learning Programming}. In \bibinfo{booktitle}{\emph{Proceedings of the 50th ACM Technical Symposium on Computer Science Education}} (Minneapolis, MN, USA) \emph{(\bibinfo{series}{SIGCSE '19})}. \bibinfo{publisher}{Assoc. for Computing Machinery}, \bibinfo{address}{New York, NY, USA}, \bibinfo{pages}{84–90}.
\newblock
\showISBNx{9781450358903}
\urldef\tempurl%
\url{https://doi.org/10.1145/3287324.3287378}
\showDOI{\tempurl}


\bibitem[Li et~al\mbox{.}(2023)]%
        {li2023wrong}
\bibfield{author}{\bibinfo{person}{Tiffany~Wenting Li}, \bibinfo{person}{Silas Hsu}, \bibinfo{person}{Max Fowler}, \bibinfo{person}{Zhilin Zhang}, \bibinfo{person}{Craig Zilles}, {and} \bibinfo{person}{Karrie Karahalios}.} \bibinfo{year}{2023}\natexlab{}.
\newblock \showarticletitle{Am {I} {Wrong}, or {Is} the {Autograder} {Wrong}? {Effects} of {AI} {Grading} {Mistakes} on {Learning}}. In \bibinfo{booktitle}{\emph{Proceedings of the 2023 {ACM} {Conference} on {International} {Computing} {Education} {Research} - {Volume} 1}} \emph{(\bibinfo{series}{{ICER} '23}, Vol.~\bibinfo{volume}{1})}. \bibinfo{publisher}{Association for Computing Machinery}, \bibinfo{address}{New York, NY, USA}, \bibinfo{pages}{159--176}.
\newblock
\showISBNx{978-1-4503-9976-0}
\urldef\tempurl%
\url{https://doi.org/10.1145/3568813.3600124}
\showDOI{\tempurl}


\bibitem[Margulieux et~al\mbox{.}(2024)]%
        {margulieux_self-regulation_2024}
\bibfield{author}{\bibinfo{person}{Lauren~E. Margulieux}, \bibinfo{person}{James Prather}, \bibinfo{person}{Brent~N. Reeves}, \bibinfo{person}{Brett~A. Becker}, \bibinfo{person}{Gozde Cetin~Uzun}, \bibinfo{person}{Dastyni Loksa}, \bibinfo{person}{Juho Leinonen}, {and} \bibinfo{person}{Paul Denny}.} \bibinfo{year}{2024}\natexlab{}.
\newblock \showarticletitle{Self-{Regulation}, {Self}-{Efficacy}, and {Fear} of {Failure} {Interactions} with {How} {Novices} {Use} {LLMs} to {Solve} {Programming} {Problems}}. In \bibinfo{booktitle}{\emph{Proceedings of the 2024 on {Innovation} and {Technology} in {Computer} {Science} {Education} {V}. 1}} \emph{(\bibinfo{series}{{ITiCSE} 2024})}. \bibinfo{publisher}{ACM}, \bibinfo{address}{New York, NY, USA}, \bibinfo{pages}{276--282}.
\newblock
\showISBNx{9798400706004}
\urldef\tempurl%
\url{https://doi.org/10.1145/3649217.3653621}
\showDOI{\tempurl}


\bibitem[Murphy et~al\mbox{.}(2012)]%
        {murphy2012explain}
\bibfield{author}{\bibinfo{person}{Laurie Murphy}, \bibinfo{person}{Ren\'{e}e McCauley}, {and} \bibinfo{person}{Sue Fitzgerald}.} \bibinfo{year}{2012}\natexlab{}.
\newblock \showarticletitle{`Explain in plain English' questions: implications for teaching}. In \bibinfo{booktitle}{\emph{Proceedings of the 43rd ACM Technical Symposium on Computer Science Education}} (Raleigh, North Carolina, USA) \emph{(\bibinfo{series}{SIGCSE '12})}. \bibinfo{publisher}{Association for Computing Machinery}, \bibinfo{address}{New York, NY, USA}, \bibinfo{pages}{385–390}.
\newblock
\showISBNx{9781450310987}
\urldef\tempurl%
\url{https://doi.org/10.1145/2157136.2157249}
\showDOI{\tempurl}


\bibitem[Nguyen et~al\mbox{.}(2024)]%
        {nguyen2024beginning}
\bibfield{author}{\bibinfo{person}{Sydney Nguyen}, \bibinfo{person}{Hannah~McLean Babe}, \bibinfo{person}{Yangtian Zi}, \bibinfo{person}{Arjun Guha}, \bibinfo{person}{Carolyn~Jane Anderson}, {and} \bibinfo{person}{Molly~Q Feldman}.} \bibinfo{year}{2024}\natexlab{}.
\newblock \showarticletitle{How {Beginning} {Programmers} and {Code} {LLMs} ({Mis})read {Each} {Other}}. In \bibinfo{booktitle}{\emph{Proceedings of the {CHI} {Conference} on {Human} {Factors} in {Computing} {Systems}}} \emph{(\bibinfo{series}{{CHI} '24})}. \bibinfo{publisher}{Association for Computing Machinery}, \bibinfo{address}{New York, NY, USA}, \bibinfo{pages}{1--26}.
\newblock
\showISBNx{9798400703300}
\urldef\tempurl%
\url{https://doi.org/10.1145/3613904.3642706}
\showDOI{\tempurl}


\bibitem[Porter and Zingaro(2024)]%
        {porter_learn_2024}
\bibfield{author}{\bibinfo{person}{Leo Porter} {and} \bibinfo{person}{Daniel Zingaro}.} \bibinfo{year}{2024}\natexlab{}.
\newblock \bibinfo{booktitle}{\emph{Learn {AI}-assisted {Python} programming: with {GitHub} {Copilot} and {ChatGPT}} (\bibinfo{edition}{first edition} ed.)}.
\newblock \bibinfo{publisher}{Manning Publications}, \bibinfo{address}{Shelter Island, New York}.
\newblock
\showISBNx{978-1-63343-778-4}


\bibitem[Prather et~al\mbox{.}(2023)]%
        {prather2023robots}
\bibfield{author}{\bibinfo{person}{James Prather}, \bibinfo{person}{Paul Denny}, \bibinfo{person}{Juho Leinonen}, \bibinfo{person}{Brett~A. Becker}, \bibinfo{person}{Ibrahim Albluwi}, \bibinfo{person}{Michelle Craig}, \bibinfo{person}{Hieke Keuning}, \bibinfo{person}{Natalie Kiesler}, \bibinfo{person}{Tobias Kohn}, \bibinfo{person}{Andrew Luxton-Reilly}, \bibinfo{person}{Stephen MacNeil}, \bibinfo{person}{Andrew Petersen}, \bibinfo{person}{Raymond Pettit}, \bibinfo{person}{Brent~N. Reeves}, {and} \bibinfo{person}{Jaromir Savelka}.} \bibinfo{year}{2023}\natexlab{}.
\newblock \showarticletitle{The {Robots} {Are} {Here}: {Navigating} the {Generative} {AI} {Revolution} in {Computing} {Education}}. In \bibinfo{booktitle}{\emph{Proceedings of the 2023 {Working} {Group} {Reports} on {Innovation} and {Technology} in {Computer} {Science} {Education}}} \emph{(\bibinfo{series}{{ITiCSE}-{WGR} '23})}. \bibinfo{publisher}{Association for Computing Machinery}, \bibinfo{address}{New York, NY, USA}, \bibinfo{pages}{108--159}.
\newblock
\showISBNx{9798400704055}
\urldef\tempurl%
\url{https://doi.org/10.1145/3623762.3633499}
\showDOI{\tempurl}


\bibitem[Prather et~al\mbox{.}(2024a)]%
        {prather2024interactions}
\bibfield{author}{\bibinfo{person}{James Prather}, \bibinfo{person}{Paul Denny}, \bibinfo{person}{Juho Leinonen}, \bibinfo{person}{David~H Smith~IV}, \bibinfo{person}{Brent~N Reeves}, \bibinfo{person}{Stephen MacNeil}, \bibinfo{person}{Brett~A Becker}, \bibinfo{person}{Andrew Luxton-Reilly}, \bibinfo{person}{Thezyrie Amarouche}, {and} \bibinfo{person}{Bailey Kimmel}.} \bibinfo{year}{2024}\natexlab{a}.
\newblock \showarticletitle{Interactions with Prompt Problems: A New Way to Teach Programming with Large Language Models}.
\newblock \bibinfo{journal}{\emph{arXiv preprint arXiv:2401.10759}} (\bibinfo{year}{2024}), \bibinfo{numpages}{30}~pages.
\newblock
\urldef\tempurl%
\url{https://doi.org/10.48550/arXiv.2401.10759}
\showURL{%
\tempurl}


\bibitem[Prather et~al\mbox{.}(2024b)]%
        {prather2024widening}
\bibfield{author}{\bibinfo{person}{James Prather}, \bibinfo{person}{Brent~N Reeves}, \bibinfo{person}{Juho Leinonen}, \bibinfo{person}{Stephen MacNeil}, \bibinfo{person}{Arisoa~S Randrianasolo}, \bibinfo{person}{Brett~A. Becker}, \bibinfo{person}{Bailey Kimmel}, \bibinfo{person}{Jared Wright}, {and} \bibinfo{person}{Ben Briggs}.} \bibinfo{year}{2024}\natexlab{b}.
\newblock \showarticletitle{The {Widening} {Gap}: {The} {Benefits} and {Harms} of {Generative} {AI} for {Novice} {Programmers}}. In \bibinfo{booktitle}{\emph{Proceedings of the 2024 {ACM} {Conference} on {International} {Computing} {Education} {Research} - {Volume} 1}} \emph{(\bibinfo{series}{{ICER} '24}, Vol.~\bibinfo{volume}{1})}. \bibinfo{publisher}{Association for Computing Machinery}, \bibinfo{address}{New York, NY, USA}, \bibinfo{pages}{469--486}.
\newblock
\showISBNx{9798400704758}
\urldef\tempurl%
\url{https://doi.org/10.1145/3632620.3671116}
\showDOI{\tempurl}


\bibitem[Sarsa et~al\mbox{.}(2022)]%
        {sarsa2022automatic}
\bibfield{author}{\bibinfo{person}{Sami Sarsa}, \bibinfo{person}{Paul Denny}, \bibinfo{person}{Arto Hellas}, {and} \bibinfo{person}{Juho Leinonen}.} \bibinfo{year}{2022}\natexlab{}.
\newblock \showarticletitle{Automatic Generation of Programming Exercises and Code Explanations Using Large Language Models}. In \bibinfo{booktitle}{\emph{Proceedings of the 2022 ACM Conference on International Computing Education Research - Volume 1}} (Lugano and Virtual Event, Switzerland) \emph{(\bibinfo{series}{ICER '22})}. \bibinfo{publisher}{Association for Computing Machinery}, \bibinfo{address}{New York, NY, USA}, \bibinfo{pages}{27–43}.
\newblock
\showISBNx{9781450391948}
\urldef\tempurl%
\url{https://doi.org/10.1145/3501385.3543957}
\showDOI{\tempurl}


\bibitem[Smith and Zilles(2024)]%
        {smith_code_2024}
\bibfield{author}{\bibinfo{person}{David~H. Smith} {and} \bibinfo{person}{Craig Zilles}.} \bibinfo{year}{2024}\natexlab{}.
\newblock \showarticletitle{Code {Generation} {Based} {Grading}: {Evaluating} an {Auto}-grading {Mechanism} for "{Explain}-in-{Plain}-{English}" {Questions}}. In \bibinfo{booktitle}{\emph{Proceedings of the 2024 on {Innovation} and {Technology} in {Computer} {Science} {Education} {V}. 1}} \emph{(\bibinfo{series}{{ITiCSE} 2024})}. \bibinfo{publisher}{Association for Computing Machinery}, \bibinfo{address}{New York, NY, USA}, \bibinfo{pages}{171--177}.
\newblock
\showISBNx{9798400706004}
\urldef\tempurl%
\url{https://doi.org/10.1145/3649217.3653582}
\showDOI{\tempurl}


\bibitem[West et~al\mbox{.}(2015)]%
        {west2015prairielearn}
\bibfield{author}{\bibinfo{person}{Matthew West}, \bibinfo{person}{Geoffrey~L. Herman}, {and} \bibinfo{person}{Craig Zilles}.} \bibinfo{year}{2015}\natexlab{}.
\newblock \showarticletitle{{PrairieLearn}: {Mastery}-based {Online} {Problem} {Solving} with {Adaptive} {Scoring} and {Recommendations} {Driven} by {Machine} {Learning}}. In \bibinfo{booktitle}{\emph{2015 {ASEE} {Annual} {Conference} \& {Exposition}}}. \bibinfo{publisher}{American Society for Engineering Education (ASEE)}, \bibinfo{address}{Seattle, WA, USA}, \bibinfo{pages}{26.1238.1--26.1238.14}.
\newblock
\showISBNx{978-0-692-50180-1}
\urldef\tempurl%
\url{https://doi.org/10.18260/p.24575}
\showDOI{\tempurl}
\newblock
\shownote{ISSN: 2153-5965}.


\bibitem[Whalley et~al\mbox{.}(2006)]%
        {whalley_australasian_2006}
\bibfield{author}{\bibinfo{person}{J. Whalley}, \bibinfo{person}{R. Lister}, \bibinfo{person}{E. Thompson}, \bibinfo{person}{T. Clear}, \bibinfo{person}{P. Robbins}, \bibinfo{person}{P.~K.~A. Kumar}, {and} \bibinfo{person}{C. Prasad}.} \bibinfo{year}{2006}\natexlab{}.
\newblock \showarticletitle{An {Australasian} {Study} of {Reading} and {Comprehension} {Skills} in {Novice} {Programmers}, {Using} the {Bloom} and {SOLO} {Taxonomies}}. In \bibinfo{booktitle}{\emph{Proceedings of the 8th {Australasian} {Conference} on {Computing} {Education}}}, \bibfield{editor}{\bibinfo{person}{Denise Tolhurst} {and} \bibinfo{person}{Samuel Mann}} (Eds.), Vol.~\bibinfo{volume}{52}. \bibinfo{publisher}{Australian Computer Society, Inc.}, \bibinfo{address}{Hobart, Australia}, \bibinfo{pages}{243--252}.
\newblock
\showISBNx{978-1-920682-34-7}
\urldef\tempurl%
\url{https://hdl.handle.net/10292/15405}
\showURL{%
\tempurl}


\end{thebibliography}

\end{document}